\documentclass[twocolumn]{autart}    

\usepackage{graphicx}          
\usepackage{xcolor}
\usepackage{amsmath}
\usepackage{array}
\usepackage{amssymb}
\usepackage{graphicx}
\usepackage[english]{babel}
\usepackage{algorithm}
\usepackage{algpseudocode}
\usepackage{xcolor}
\usepackage{caption}
\usepackage{cuted}
\usepackage{arydshln}
\usepackage{pmat}
\usepackage[utf8]{inputenc}
\usepackage{apptools}
\AtAppendix{\counterwithin{lemapp}{section}}
\usepackage[titletoc]{appendix}
\newtheorem{lemapp}{Lemma}

\newcounter{todocounter}
\usepackage[prependcaption,colorinlistoftodos]{todonotes}

\let\emptyset\varnothing

\newcommand{\R}{\mathbb{R}}
\newcommand{\C}{\mathbb{C}}
\newcommand{\PP}{\mathbb{P}}
\newcommand{\MM}{\mathbb{M}}
\newcommand{\N}{\mathbb{N}}

\newcommand{\calH}{\mathcal{H}}
\newcommand{\calL}{\mathcal{L}}

\DeclareMathOperator{\rank}{rank}
\DeclareMathOperator{\leftker}{leftker}

\DeclareMathOperator{\Real}{\mathrm{Re}}
\DeclareMathOperator{\Imag}{\mathrm{Im}}
\DeclareMathOperator{\rowspace}{\mathrm{rowspace}}

\newtheorem{problem}{Problem}
\usepackage{tikz}
\usepackage{graphicx}      
\usepackage{pgfplots} 
\pgfplotsset{compat=newest} 
\pgfplotsset{plot coordinates/math parser=false}
\newlength\figureheight
\newlength\figurewidth
\usepackage{wrapfig}
\usepackage[american]{circuitikz}
\makeatletter
\ctikzset{bipoles/vsources/minus/sign/rotation/.initial=0}
\pgfcircdeclarebipole{}{\ctikzvalof{bipoles/vsourceam/height}}{vsourceAM}{\ctikzvalof{bipoles/vsourceam/height}}{\ctikzvalof{bipoles/vsourceam/width}}{
	\pgfsetlinewidth{\pgfkeysvalueof{/tikz/circuitikz/bipoles/thickness}\pgfstartlinewidth}
	\pgfpathellipse{\pgfpointorigin}{\pgfpoint{0}{\pgf@circ@res@up}}{\pgfpoint{\pgf@circ@res@left}{0}}
	\pgfusepath{draw}
	\ifpgf@circ@oldvoltagedirection
	\pgftext[bottom,rotate=90,y=\ctikzvalof{bipoles/vsourceam/margin}\pgf@circ@res@down]{$+$}
	\pgftext[top,rotate=90,y=\ctikzvalof{bipoles/vsourceam/margin}\pgf@circ@res@up]{$-$}
	\else
	\pgftext[bottom,rotate=90,y=\ctikzvalof{bipoles/vsourceam/margin}\pgf@circ@res@down]{\rotatebox{\ctikzvalof{bipoles/vsources/minus/sign/rotation}}{$-$}}
	\pgftext[top,rotate=90,y=\ctikzvalof{bipoles/vsourceam/margin}\pgf@circ@res@up]{$+$}
	\fi
}

\pgfcircdeclarebipole{}{\ctikzvalof{bipoles/cvsourceam/height}}{cvsourceAM}{\ctikzvalof{bipoles/cvsourceam/height}}{\ctikzvalof{bipoles/cvsourceam/width}}{
	\pgfsetlinewidth{\pgfkeysvalueof{/tikz/circuitikz/bipoles/thickness}\pgfstartlinewidth}
	\pgfpathmoveto{\pgfpoint{\pgf@circ@res@left}{\pgf@circ@res@zero}}
	\pgfpathlineto{\pgfpoint{\pgf@circ@res@zero}{\pgf@circ@res@up}}
	\pgfpathlineto{\pgfpoint{\pgf@circ@res@right}{\pgf@circ@res@zero}}
	\pgfpathlineto{\pgfpoint{\pgf@circ@res@zero}{\pgf@circ@res@down}}
	\pgfpathlineto{\pgfpoint{\pgf@circ@res@left}{\pgf@circ@res@zero}}
	\pgfusepath{draw}
	
	\ifpgf@circ@oldvoltagedirection
	\pgftext[bottom,rotate=90,y=\ctikzvalof{bipoles/cvsourceam/margin}\pgf@circ@res@left]{$+$}
	\pgftext[top,rotate=90,y=\ctikzvalof{bipoles/cvsourceam/margin}\pgf@circ@res@right]{$-$}
	\else
	\pgftext[bottom,rotate=90,y=\ctikzvalof{bipoles/cvsourceam/margin}\pgf@circ@res@left]{\rotatebox{\ctikzvalof{bipoles/vsources/minus/sign/rotation}}{$-$}}
	\pgftext[top,rotate=90,y=\ctikzvalof{bipoles/cvsourceam/margin}\pgf@circ@res@right]{$+$}
	\fi
}

\makeatletter
\def\addlegendimage{\pgfplots@addlegendimage}

\makeatother

\begin{document}

\begin{frontmatter}

\title{From data to reduced-order models via moment matching\thanksref{footnoteinfo}} 

\thanks[footnoteinfo]{The material in this paper was partially presented at the 21st IFAC World Congress, July 11-17, 2020, Berlin, Germany. Corresponding author A.~M.~Burohman.}

\author[First,Second,Third]{Azka M. Burohman} 
\author[First,Second]{Bart Besselink} 
\author[First,Third]{Jacquelien M. A. Scherpen}
\author[First,Second]{M. Kanat Camlibel}

\address[First]{Jan C. Willems Center for Systems and Control, University of Groningen, The~Netherlands.}
\address[Second]{Bernoulli Institute for Mathematics, Computer Science and Artificial Intelligence, University~of~Groningen, The~Netherlands. (e-mail:\{a.m.burohman, b.besselink, m.k.camlibel\}@rug.nl)}
\address[Third]{Engineering and Technology Institute Groningen, University of Groningen, The Netherlands. (e-mail: j.m.a.scherpen@rug.nl)}


\begin{abstract}                          
A new method for data-driven interpolatory model reduction is presented in this paper. Using the so-called data informativity perspective, we define a framework that enables the computation of moments at given (possibly complex) interpolation points based on time-domain input-output data only, without explicitly identifying the high-order system. Instead, by characterizing the set of all systems explaining the data, necessary and sufficient conditions are provided under which all systems in this set share the same moment at a given interpolation point. Moreover, these conditions allow for explicitly computing these moments. Reduced-order models are then derived by employing a variation of the classical rational interpolation method. The condition to enforce moment matching model reduction with prescribed poles is also discussed as a means to obtain stable reduced-order models. An example of an electrical circuit illustrates this framework.
\end{abstract}

\end{frontmatter}

\section{Introduction}
In the modeling of complex phenomena, e.g., in modern engineering systems, high-order dynamical systems naturally appear. They result from either the inherent complexity of (engineering or physical)  systems or from the discretization of partial differential equations. The approximation of the input-output behavior of such a model by a model of lower order is known as model reduction, which has become a crucial tool in the analysis and control of complex systems. Several model reduction techniques for linear systems have been developed in the systems and control community, of which a thorough exposition can be found in \cite{antoulas2005approximation}. The most popular techniques can roughly be divided into two main groups. The first group is based on energy functions characterized by Gramians and contains balancing methods \cite{benner2011lyapunov,4047847,moore1981principal,1102945} and optimal Hankel norm approximation \cite{glover1984all}. The other contains interpolatory or Krylov projection-based methods and this paper belongs to this second group. 

Interpolatory techniques form a popular class of model reduction approaches, since they are numerically stable and, therefore, applicable to models of very large order. These methods are aimed at constructing a reduced-order model whose transfer function interpolates that of the original high-order model at selected interpolation points, e.g., \cite{doi:10.1137/1.9781611976083}. Moment matching techniques form an example of interpolatory methods and were originally developed in the field of numerical mathematics, see, e.g., \cite{feldmann1995efficient,grimme1997krylov}. By exploiting Krylov subspaces and projection, these methods achieve interpolation without explicitly evaluating the transfer function. Additionally, these methods are well suited for interpolating also the (higher-order) derivatives of the transfer function known as \emph{moments}, see also \cite{gallivan2004model,gugercin2008h_2}. 
A time-domain perspective on moment matching is given in \cite{astolfi2010model}, building on a relation between projection matrices for moment matching and Sylvester equations \cite{gallivan2002generality,gallivan2004sylvester}. This time-domain perspective enables extensions of moment matching to more general system classes, see \cite{scarciotti2017nonlinear}.

The majority of the existing model reduction methods rely on the availability of a state-space model or transfer function of the system to be reduced. In this paper, however, we develop a \emph{data-driven} moment matching method, i.e., based exclusively on time-domain {noise free} input-output {data} obtained from the high-order system. This is motivated by the observation that, in many cases, an explicit system model (or access to it) is not available {while one can access input-output data. Examples include high-fidelity simulation models in engineering, for instance fluid flow and structural mechanics (see e.g. \cite{rowley_2005}).} 

A data-driven model reduction approach by moment matching has been introduced in \cite[Chapter~4]{doi:10.1137/1.9781611976083}. Relying on \emph{frequency-domain} data, this so-called Loewner framework has strong connections to classical rational interpolation,
see \cite{anderson1990rational,antoulas1990minimal}. Rational interpolation has been studied extensively and allows for obtaining systems that, in addition to achieving interpolation of frequency-domain data, guarantee further system properties such as a minimal McMillan degree or stability \cite{antoulas1989problem,antoulas1990solution,ball1994stability}. Taking a model reduction perspective, the desired interpolation conditions might result from a higher-order system, in which case the Loewner framework allows for obtaining reduced-order systems that achieve interpolation as well as further properties, see \cite{antoulas2016loewner,mayo2007framework}. Specifically, the preservation of stability \cite{gosea2016stability} and passivity  \cite{antoulas2005new} is considered, as is optimal approximation in the $\calH_2$ system norm, see \cite{beattie2012realization}. To enable the use of time-domain data (rather than frequency-domain data) in this framework, \cite{peherstorfer2017data} estimates transfer function values at given interpolation points by exploiting the relation between time- and frequency-domain data via the (discrete) Fourier transform, after which standard interpolatory methods can be used.

Data-driven model reduction from given \emph{time-domain} data has also become an attractive topic in recent years. Belonging to the class of moment matching methods, algorithms for computing (a least-square approximation of) moments of linear or nonlinear systems are proposed in \cite{scarciotti2017data}, building on the framework of \cite{astolfi2010model}. These (estimated) moments are then used to construct families of reduced-order models. This methods however relies on specifically chosen input data to guarantee that the resulting data (obtained from a steady-state response) is suitable for estimating a moment. In the class of Gramian-based methods, and in case the input data are assumed to be persistently exciting, a balanced realization of the high-order system generating the input-output data can be obtained using algorithms proposed in \cite{markovsky2005algorithms}. A behavioral approach is used in \cite{rapisarda2011identification} to obtain so-called lossless or bounded-real balanced realizations. 
After obtaining such balanced realization, the reduced-order model can readily be computed. In another approach, if measurements on the state trajectories are available, dynamic mode decomposition (DMD) provides a way to find a linear model that best fits the given trajectories in the $\calL_2$ sense as in \cite{proctor2016dynamic}. In \cite{monshizadeh2020amidst}, if the state trajectories are available, a lower-order model expressed in terms of data is uncovered such that its complexity matches that of the given data. A data-driven model reduction approach for a class of nonlinear systems is presented in \cite{kawano2019data}.

In contrast to existing works, this paper presents a method for obtaining moments at a given interpolation point \emph{directly from time-domain input-output data}, without requiring that the data are sufficiently rich to uniquely identify the high-order system (that generates the data). In particular, the contributions of this paper are threefold.



First, we introduce a new framework for computing moments from data, without necessarily identifying the full model. This perspective builds on the concept of data informativity introduced in \cite{van2019data}, which aims to find conditions on a given data set such that specific system properties can be concluded from them. This perspective has been applied successfully in the scope of system analysis \cite{eising2020data,van2019data} and control design \cite{trentelman2020informativity,van2020noisy}.
We stress that this framework does not require that the (high-order) system can be fully identified, as would for example be implied by the availability of impulse response data \cite{ho1966effective} or by  persistently exciting data, see the so-called fundamental lemma in \cite{willems2005note}. The importance of such persistently exciting data is well-known in the field of system identification \cite{Ljung1999System,verhaegen2007filtering}.
{Contrary to system identification, we take a data informativity perspective. Specifically, we define the concept of \emph{data informativity for interpolation} at some interpolation point as a notion that guarantees that the data are rich enough to uniquely determine the value of the transfer function of the high-order system at that interpolation point.} We then generalize this concept to higher-order moments.


Second, we provide necessary and sufficient conditions for given input-output data to be informative for interpolation and/or moment matching of order $k$ at a given interpolation point. We show that these conditions are weaker than those for identification, such that there is no need to fully identify the higher-order system. These conditions also lead to an approach for the \emph{computation} of the moments on the basis of data. We stress that this approach yields the \emph{exact} moments, instead of an approximation. 
  
 Finally, we construct the reduced-order model based on the computed moments. Our approach has a natural connection to classical rational interpolation. Therefore, the computation of the reduced-order model in this paper is similar to the rational interpolation method presented in \cite{antoulas1990minimal}. In addition, inspired by \cite{astolfi2010model}, we show that our method enables to obtain a reduced-order model with prescribed poles, which enables the construction of stable reduced-order models.

The remainder of this paper is organized as follows. First, a detailed problem setting is given in Section~\ref{sectionII}. In Section~\ref{Section_inf_intp}, data informativity for interpolation is introduced, which includes an illustrative example. Next, informativity for higher-order moments and the computation of them are presented in Section~\ref{Section_higher}. The computed moments are then exploited to compute a reduced-order model in Section~\ref{Section_rom}, which includes the discussion on reduction with prescribed poles. Finally, conclusions are stated in Section~\ref{Section_conclusion}.




\section{Problem formulation}\label{sectionII}
In this section, we will introduce a framework to deal with data-driven model reduction via interpolation and moment matching.

Consider a discrete-time single-input single-output{\footnote{{A similar consideration can be performed for multi-input multi-output systems using tangential interpolation (see e.g. \cite{gallivan2004model}).}}} system of the form
\begin{equation}\label{Eqn_true_sys}
\begin{split}
&y_{t+n} + \bar{p}_{n-1}y_{t+n-1} + \cdots + \bar{p}_1 y _{t+1} + \bar{p}_0y_t \\ 
&\quad = \bar{q}_nu_{t+n} + \bar{q}_{n-1}u_{t+n-1} +  \cdots + \bar{q}_1u_{t+1} + \bar{q}_0u_t,
\end{split}
\end{equation}
where $u$ denotes the scalar input, $y$ the scalar output and $t \in \N$ the discrete time. We denote the parameters of (\ref{Eqn_true_sys}) by
$
\bar{p}=\begin{bmatrix}
\bar{p}_0 & \bar{p}_1 & \cdots & \bar{p}_{n-1}
\end{bmatrix}
\quad \text{and} \quad 
\bar{q} =\begin{bmatrix}
\bar{q}_0 & \bar{q}_1 & \cdots & \bar{q}_n
\end{bmatrix}.
$%
Throughout the paper, we assume that the parameters $\begin{bmatrix}
\bar{q} & -\bar{p}
\end{bmatrix}$ are \emph{unknown} but $n\geq 0$ is \emph{known} and we have access to \emph{input-output data} given by  
\begin{equation*}
U= \begin{bmatrix}
\bar{u}_0 & \bar{u}_1 & \cdots & \bar{u}_T
\end{bmatrix} \quad \text{and} \quad Y= \begin{bmatrix}
\bar{y}_0 & \bar{y}_1 & \cdots & \bar{y}_T
\end{bmatrix}%
\end{equation*}
that are generated by the `true' system (\ref{Eqn_true_sys}), i.e.,
\begin{equation}\label{sys_true_data}
\begin{split}
&\bar{y}_{t+n} + \bar{p}_{n-1}\bar{y}_{t+n-1} + \cdots + \bar{p}_1 \bar{y} _{t+1} + \bar{p}_0\bar{y}_t \\ 
&\quad = \bar{q}_n\bar{u}_{t+n} + \bar{q}_{n-1}\bar{u}_{t+n-1} +  \cdots + \bar{q}_1\bar{u}_{t+1} + \bar{q}_0\bar{u}_t,
\end{split}%
\end{equation}
for $t=0,1,\ldots,T-n$ and some $T\geq n$.

We do \emph{not} assume that the data $(U,Y)$ uniquely identify the system (\ref{Eqn_true_sys}). To determine all systems that are compatible with the data $(U,Y)$, consider an input-output system of the form
\begin{equation}\label{Eqn_datasys}
\begin{split}
&{y}_{t+n} + {p}_{n-1}{y}_{t+n-1} + \cdots + {p}_1 {y}_{t+1} + {p}_0{y}_t \\ 
&\quad = {q}_n{u}_{t+n} + {q}_{n-1}{u}_{t+n-1} +  \cdots + {q}_1{u}_{t+1} + {q}_0{u}_t.
\end{split}
\end{equation}
The data $(U,Y)$ can be generated by this system if and only if 
\begin{equation*}
\begin{split}
&\bar{y}_{t+n} + {p}_{n-1}\bar{y}_{t+n-1} + \cdots + {p}_1 \bar{y}_{t+1} + {p}_0\bar{y}_t \\ 
&\quad = {q}_n\bar{u}_{t+n} + {q}_{n-1}\bar{u}_{t+n-1} +  \cdots + {q}_1\bar{u}_{t+1} + {q}_0\bar{u}_t
\end{split}
\end{equation*}
for $t=0,1,\ldots,T-n$. This system of linear equations can be written more compactly as 
\begin{equation}\label{linear1}
\begin{bmatrix}
q & -p
\end{bmatrix} \begin{bmatrix}
H_{n}(U) \\ \bar{H}_{n}(Y)
\end{bmatrix} = \begin{bmatrix}
\bar{y}_n & \cdots & \bar{y}_{T}
\end{bmatrix},
\end{equation}
where 
$
{p}=\begin{bmatrix}
p_0 & p_1 & \cdots & p_{n-1}
\end{bmatrix}, \quad 
{q}=\begin{bmatrix}
q_0 & q_1 & \cdots & q_n
\end{bmatrix},
$
and $H_\ell(U)$ denotes the Hankel matrix of depth $\ell$ obtained from $U$, i.e.,
\begin{equation*} 
H_{\ell}(U) = \begin{bmatrix}
\bar{u}_0  &  \bar{u}_1  & \cdots & \bar{u}_{T-\ell} \\
\bar{u}_1  & \bar{u}_2 &  \cdots  &  \bar{u}_{T-\ell+1} \\
\vdots  & \!\vdots &  \ddots  & \vdots \\
\bar{u}_{\ell}  & \bar{u}_{\ell+1} & \cdots &  \bar{u}_{T}
\end{bmatrix}  \in  \R^{(\ell+1)\times (T-\ell+1)}.
\end{equation*}
The Hankel matrix $H_\ell(Y)$ is defined in a similar fashion and, finally, $\bar{H}_{\ell}(Y)$ is obtained from $H_\ell(Y)$ by deleting its last row. Now, we can identify the set of all systems that can generate the data $(U,Y)$ with the set 
\begin{equation}\label{Set_UY}
\Sigma_{U,Y}=\left\lbrace  \left. \begin{bmatrix}
q & -p
\end{bmatrix} \in \R^{1 \times (2n+1)}\right| \quad (\ref{linear1}) \quad  \text{holds} \right\rbrace.
\end{equation}
Since the data $(U,Y)$ are generated by the `true' system (\ref{Eqn_true_sys}), i.e., (\ref{sys_true_data})  is satisfied, we clearly have that $\begin{bmatrix}
\bar{q} & -\bar{p}
\end{bmatrix} \in \Sigma_{U,Y}$. An obvious question to ask is when the data uniquely determine the `true' system.
\begin{defn}\label{Def_sysID}
	The data $(U,Y)$ are \emph{informative for system identification} if $\Sigma_{U,Y}$ is a singleton.
\end{defn}
Data informativity for system identification can easily be characterized as follows.
\begin{prop}
	The data $(U,Y)$ are informative for system identification if and only if 
	\begin{equation}\label{Cond_sysID}
	\rank\begin{bmatrix}
	H_n(U) \\ \bar{H}_n(Y)
	\end{bmatrix}=\rank\begin{bmatrix}
	H_n(U) \\H_n(Y)
	\end{bmatrix}=2n+1.
	\end{equation}	
\end{prop}
\begin{pf}
	It is obvious that the first and the second equality in (\ref{Cond_sysID}) are equivalent to the existence and uniqueness of the solution of  (\ref{linear1}), respectively. 
	\hfill\hfill  \qed
\end{pf}
\begin{rem}
	In system identification, sufficient conditions for identification are typically stated in terms of the so-called \emph{persistency of excitation} condition on the \emph{input data only}. Here,  the data $U$ are persistently exciting of order $\ell + 1$ if there exists an integer $T$ such that $H_{\ell}(U)$ has full rank $\ell + 1$. Then, a sufficient condition for system identification (i.e., uniquely constructing a transfer function from data) is that $U$ is persistently exciting of order $2n+1$, see \cite{verhaegen2007filtering,willems2005note}. {It can be checked by \cite[Lemma~9.1]{verhaegen2007filtering} and \cite[Corollary~2]{willems2005note}
		that if $n$ is known, then this persistency of excitation condition implies (\ref{Cond_sysID})}. We stress however that (\ref{Cond_sysID}) is \emph{necessary and sufficient} for system identification.
\end{rem}

\subsection{Moments of a discrete-time linear system}\label{subsectionIIB}
Next, we will introduce moments of the system (\ref{Eqn_datasys}). To do so, we first introduce the forward shift operator $z$ defined as $zf_t = f_{t+1}$. Then, we rewrite (\ref{Eqn_datasys}) in the form
\begin{equation}\label{Eqn_polyyu}
\begin{split}
&\left(z^n+p_{n-1}z^{n-1}+\cdots+p_1z+p_0\right)y_t \\ & \quad \quad \quad  = \left(q_nz^n+q_{n-1}z^{n-1}+\cdots+q_1z+q_0\right)u_t.
\end{split}
\end{equation}
We denote the polynomial on the left- and right-hand sides of (\ref{Eqn_polyyu}), respectively, as
\begin{equation}\label{nota_PQ}
\begin{split}
&P(z)=z^n+p_{n-1}z^{n-1}+\cdots+p_1z+p_0, \\
&Q(z)=q_nz^n+q_{n-1}z^{n-1}+\cdots+q_1z+q_0.
\end{split}
\end{equation}
Now, we are in a position to define the $0$-th moment for the system (\ref{Eqn_datasys}). 
\begin{defn}[$0$-th moment]\label{Def_0moment}
	Given an interpolation point $\sigma \in \C$, a number $M_0 \in \C$ is said to be a \emph{$0$-th moment at $\sigma$} of the discrete-time system (\ref{Eqn_datasys}) if 
	\begin{equation}\label{defmoment}
	P(\sigma)M_0 = Q(\sigma).
	\end{equation}
	In this case, we also write $M_0=M_0(\sigma)$.
\end{defn}
\begin{rem}\label{Remark_M0}
	For a discrete-time system (\ref{Eqn_datasys}) with transfer function $G(z)=Q(z)/P(z)$,
	the $0$-th moment at $\sigma$ is typically defined as the complex number $Q(\sigma)/P(\sigma)$.
	This, however, requires that $P(\sigma)~\neq~0$. In other words, the $0$-th moment is not defined when $\sigma$ is a pole of $G(z)$. The notion in Definition~\ref{Def_0moment} is a slight generalization as it allows to define a moment in case both $P(\sigma)=0$ and $Q(\sigma)=0$, i.e., there is a pole-zero cancellation at $\sigma$. We stress that minimality is not assumed for (\ref{Eqn_datasys}). Therefore, \emph{any} complex number $M_0$ is regarded as a $0$-th moment at $\sigma$ by Definition~\ref{Def_0moment} in case $P(\sigma)=Q(\sigma)=0$. 
\end{rem}


Given (\ref{nota_PQ}), condition (\ref{defmoment}) can be written as the linear equation 
\begin{equation}\label{linear2}
\begin{bmatrix}
q & -p
\end{bmatrix}\begin{bmatrix}
\gamma_n(\sigma) \\ M_0 \gamma_{n-1}(\sigma)
\end{bmatrix} = M_0\sigma^n,
\end{equation}
where
$
\gamma_{\ell}(z) = \begin{bmatrix}
1 & z & \cdots & z^{\ell}
\end{bmatrix}^T.
$
Then, the expression (\ref{linear2}) allows for defining 
\begin{equation}\label{Set_sigmaM}
\Sigma^0_{\sigma,M_0}=\left\{ \left.\begin{bmatrix}
q & -p
\end{bmatrix}\in \R^{1 \times (2n+1)} \right| \quad (\ref{linear2}) \quad  \text{holds} \right\}
\end{equation}
as the set of all (parameters of) systems of order $n$ that have $0$-th moment $M_0$ at $\sigma$. 


Let $f^{(j)}$ denote the $j$-th derivative of $f$, i.e., $f^{(j)}(z)=\frac{d^j}{dz^j}f(z)$.
Next, we define higher order moments in a recursive manner.
\begin{defn}[$k$-th moment] \label{Def_kthmoment}
	Given an interpolation point $\sigma \in \C$ and $j$-th moments (at $\sigma$) $M_j$ for $j=0,1,\ldots, k-1$. Then, a number $M_k \in \C$ is said to be a \emph{$k$-th moment at $\sigma$} of the discrete-time system (\ref{Eqn_datasys}) if 
	\begin{equation}\label{momentqp}
	Q^{(k)}(\sigma) = \sum_{j=0}^{k}  \binom{k}{j}M_j P^{(k-j)}(\sigma),
	\end{equation}
	where 
	$\binom{k}{j}= \frac{k!}{j!(k-j)!}$
	is the binomial coefficient. In this case, we also write $M_k=M_k(\sigma)$. 
\end{defn}


\begin{rem}
	The definition of $k$-th moment in Definition~\ref{Def_kthmoment} is related to the classical definition of moments as in, e.g., \cite{antoulas2005approximation} and \cite{astolfi2010model}. Classically, the $k$-th moment of a transfer function $G(z)$ at $\sigma$ is defined by its $k$-th derivative with respect to $z$ evaluated at $z=\sigma$, i.e.,
	\begin{equation*}
	M_k(\sigma)=\left.\frac{d^k}{dz^k}G(z) \right|_{z=\sigma}, \quad k\geq 0.
	\end{equation*}
	With $G(z)=Q(z)/P(z)$, the derivatives (up to order $k$) of $G(z)$, $P(z)$ and $Q(z)$ satisfy
	\begin{equation}\label{Eqn_QGP}
	Q^{(k)}(z) = \sum_{j=0}^{k} \binom{k}{j} G^{(j)}(z)P^{(k-j)}(z).
	\end{equation}
	Evaluating (\ref{Eqn_QGP}) at $z=\sigma$ yields (\ref{momentqp}). Hence, (\ref{momentqp}) reduces to the classical definition of $k$-th moment when $P(\sigma) \neq 0$. Nonetheless, Definition \ref{Def_kthmoment} is a slight generalization for reasons similar as those outlined in Remark \ref{Remark_M0}.
\end{rem}

Similarly as in \eqref{linear2}, equation (\ref{momentqp}) can be written as a linear equation in $\begin{bmatrix}
q & -p
\end{bmatrix}$ as
\begin{equation}\label{Eqn_kthmoment}
\begin{split}
\begin{bmatrix}
q & -p
\end{bmatrix}&\begin{bmatrix}
\gamma_n^{(k)}(\sigma) \\ \sum_{j=0}^{k}  \binom{k}{j} M_ j\gamma_{n-1}^{(k-j)}(\sigma)
\end{bmatrix}\\
&=\sum_{j=0}^{k} \binom{k}{j}\frac{n!}{(n-k+j)!} M_j\sigma^{n-k+j},
\end{split}
\end{equation}
where
$\gamma_{\ell}^{(j)}(\sigma)=\left.\frac{d^j}{dz^j}\gamma_{\ell}(z) \right|_{z=\sigma}$.

We define the set of all (parameters of) systems having $M_k$ as the $k$-th moment at $\sigma$ by
\begin{equation*}
\Sigma^k_{\sigma,M_k}=\left\{ \left.\begin{bmatrix}
q & -p
\end{bmatrix} \in \R^{1 \times (2n+1)} \right| \quad  (\ref{Eqn_kthmoment}) \quad \text{holds}\right\}.
\end{equation*}
Note that these sets are also defined by a recursion, {i.e., this definition of set holds for $k=1,2,\ldots,n$.}


\subsection{Informativity problem for moment matching}
Recall that we are interested in the moments of the true system (\ref{Eqn_true_sys}), but only have the input-output data $(U,Y)$ available. Given an interpolation point $\sigma \in \C$, we are interested in finding (necessary and sufficient) conditions for the data $(U,Y)$ to be sufficiently rich to allow for computing the moments at $\sigma$. These conditions will allow for the construction of reduced-order models that achieve moment matching.

Starting with the $0$-th moment, recall that all systems with moment $M_0$ at $\sigma$ are given by the set $\Sigma^0_{\sigma,M_0}$. However, as $\begin{bmatrix}
\bar{q} & -\bar{p}
\end{bmatrix} \in \Sigma_{U,Y}$, it is sufficient to ask for $\Sigma_{U,Y} \subseteq \Sigma^0_{\sigma,M_0}$. This motivates the following definition.
\begin{defn}\label{def_inf1}
	The data $(U,Y)$ are \emph{informative for interpolation at $\sigma$} if there exists a unique $M_0$ such that $\Sigma_{U,Y} \subseteq \Sigma^0_{\sigma,M_0}$.
\end{defn}

Note that the condition $\Sigma_{U,Y} \subseteq \Sigma^0_{\sigma,M_0}$ requires all systems explained by the data to have the same $0$-th moment $M_0$ at $\sigma$. It is clear that, if the data $(U,Y)$ are informative for system identification (see Definition~\ref{Def_sysID}) and $P(\sigma) \neq 0$, then such unique $M_0$ exists. In this paper, we however look for conditions for informativity for interpolation that are strictly weaker than those for identification. This motivates the following problem statement. 
\begin{problem}\label{problem2}
	Find necessary and sufficient conditions such that $(U,Y)$ are informative for interpolation at $\sigma$. Furthermore, if the data satisfy these conditions, then find $M_0$.
\end{problem}

Once we know that the data are informative for interpolation at $\sigma$ (i.e., for matching of the $0$-th moment), we are also interested to match higher-order moments. 
\begin{defn}\label{Def_informativityMk}
	The data $(U,Y)$ are \emph{informative for moment matching of order $k$ at $\sigma$} if 
	\begin{enumerate}
		\item they are informative for moment matching of order $j$ at $\sigma$ for $j=0,1,2,\ldots, k-1$ (where moment matching of order $0$ is understood as interpolation), and 
		\item there exists a unique $M_k$ such that $\Sigma_{U,Y} \subseteq \Sigma^k_{\sigma,M_k}$.
	\end{enumerate}
\end{defn}
Note that Definition~\ref{Def_informativityMk} follows the recursive structure of the notion of $k$-th moment in Definition \ref{Def_kthmoment}. Finally, we are interested in the following problem.
\begin{problem}\label{problem3}
	Find necessary and sufficient conditions such that $(U,Y)$ are informative for moment matching of order $k$ at $\sigma$, with $k=1,2,\ldots$. Furthermore, if the data satisfy these conditions, then find $M_k$.
\end{problem}

{In general, the choice of interpolation points depends on the frequency range of interest that should be pre-specified by the user. As such, it is outside the scope of the paper.} 
\section{Data informativity for interpolation}\label{Section_inf_intp}
In this section, we will provide necessary and sufficient conditions under which the data are informative for interpolation (i.e., moment matching of order $0$) at a given interpolation point.

The following lemma will play an instrumental role in the characterization of informativity for interpolation. 
\begin{lem}\label{Thm_two_incl}
 Let $\sigma$ and $M_0$ be complex numbers. Then, the inclusion $\Sigma_{U,Y} \subseteq \Sigma^0_{\sigma,M_0}$ holds if and only if there exists $\xi \in \C^{T-n+1}$ such that
	\begin{equation}\label{Eqn_lin2}
	\begin{bmatrix}
	H_{n}(U) & 0 \\ H_{n}(Y) & -\gamma_n(\sigma)
	\end{bmatrix} \begin{bmatrix}
	\xi \\ M_0
	\end{bmatrix}=\begin{bmatrix}
	\gamma_n(\sigma) \\ 0
	\end{bmatrix}.%
	\end{equation}%
\end{lem}%
\begin{pf}	
	Note that $\Sigma_{U,Y} \subseteq \Sigma^0_{\sigma,M_0}$ is equivalent to saying that every solution of (\ref{linear1}) is also a solution of (\ref{linear2}). Note that (\ref{linear2}) can be split into a real and imaginary part. Then, since $\begin{bmatrix}
	q & -p
	\end{bmatrix}$ is real, we have by Lemma~\ref{Lemma2tr} in the appendix that $\Sigma_{U,Y} \subseteq \Sigma^0_{\sigma,M_0}$ if and only if
		\begin{equation}\label{leftker_real}
	\leftker\begin{bmatrix}
	H_n(U) \\ H_n(Y)
	\end{bmatrix} \subseteq \leftker\Real\begin{bmatrix}
	\gamma_n(\sigma) \\  M_0 \gamma_n(\sigma)
	\end{bmatrix}
	\end{equation}
	and
	\begin{equation}\label{leftker_imag}
	\leftker\begin{bmatrix}
	H_n(U) \\ H_n(Y)
	\end{bmatrix} \subseteq \leftker\Imag\begin{bmatrix}
	\gamma_n(\sigma) \\  M_0 \gamma_n(\sigma)
	\end{bmatrix},
	\end{equation}
	where $\Real$ and $\Imag$ denote, respectively, real and imaginary parts, {and $\leftker$ denotes the left kernel}. Note that (\ref{leftker_real}) and (\ref{leftker_imag}) are equivalent to the existence of real vectors $\xi_R$ and $\xi_I$ such that
		\begin{equation*}
	\begin{bmatrix}
	H_n(U) \\ H_n(Y)
	\end{bmatrix} \begin{bmatrix}\xi_R&\xi_I \end{bmatrix} = \begin{bmatrix}\Real \begin{bmatrix}
	\gamma_n(\sigma) \\ M_0 \gamma_n(\sigma)
	\end{bmatrix}& \Imag\begin{bmatrix}
	\gamma_n(\sigma) \\ M_0 \gamma_n(\sigma)
	\end{bmatrix}
	\end{bmatrix}
	\end{equation*}
By denoting $\xi=\xi_R + i \xi_I$, we can conclude that $\Sigma_{U,Y} \subseteq \Sigma^0_{\sigma,M_0}$ if and only if there exists $\xi \in \C^{T-n+1}$ such that
$	\begin{bmatrix}
	H_{n}(U)  \\ H_{n}(Y) 
	\end{bmatrix} \xi = \begin{bmatrix}
	\gamma_n(\sigma) \\ M_0 \gamma_n(\sigma) 
	\end{bmatrix}$. \qed
\end{pf}

\begin{rem}
Note that the system parameters $\begin{bmatrix}
	q & -p
	\end{bmatrix}$ are restricted to be real (see the sets (\ref{Set_UY}) and (\ref{Set_sigmaM})) whereas the interpolation points $\sigma$ might be taken complex. Hence, the main result of Lemma~\ref{Thm_two_incl} is that the (real-valued) inclusion  $\Sigma_{U,Y} \subseteq \Sigma^0_{\sigma,M_0}$ is equivalent to, the possibly complex-valued, condition (\ref{Eqn_lin2}).
\end{rem}

By using the above result, we provide the first main result of this paper: necessary and sufficient conditions for data informativity for interpolation.
\begin{thm}\label{Thm_1}
	The data $(U,Y)$ are informative for interpolation at $\sigma$ if and only if 
	\begin{equation}\label{condition2}
	\rank\!\begin{bmatrix}
	H_{n}(U) \!& \!0 \! & \!\gamma_n(\sigma) \\ H_{n}(Y) \! & \gamma_n(\sigma)\! & \!0 
	\end{bmatrix} \!= \!\rank\!\begin{bmatrix}
	H_{n}(U)\! & \!0\! \\ H_{n}(Y) \! & \gamma_n(\sigma)
	\end{bmatrix} \! \!\!%
	\end{equation}%
	and 
	\begin{equation}\label{condition3}
	\rank\begin{bmatrix}
	H_{n}(U) & 0 \\ H_{n}(Y) & \gamma_n(\sigma)
	\end{bmatrix} = \rank\begin{bmatrix}
	H_{n}(U)  \\ H_{n}(Y) 
	\end{bmatrix} +1.%
	\end{equation}%
\end{thm}
\vspace*{-8mm}
\begin{pf}
	By Definition~\ref{def_inf1}, the data $(U,Y)$ are informative for interpolation at $\sigma$ if there exists a unique $M_0$ such that $\Sigma_{U,Y} \subseteq \Sigma^0_{\sigma,M_0}$. On the one hand, the existence of such an $M_0$ is equivalent to the existence of $\xi$ such that
		\begin{equation}\label{Eqn_lin_0}
	\begin{bmatrix}
	H_{n}(U) & 0 \\ H_{n}(Y) & -\gamma_n(\sigma)
	\end{bmatrix} \begin{bmatrix}
	\xi \\ M_0
	\end{bmatrix}=\begin{bmatrix}
	\gamma_n(\sigma) \\ 0
	\end{bmatrix}
	\end{equation}
	due to Lemma~\ref{Thm_two_incl}. On the other hand, the uniqueness of $M_0$ is equivalent to the implication
	\begin{equation}\label{Eqn_uniqueness}
	\begin{split}
	\!&\begin{bmatrix}
		H_{n}(U) & 0 \\ H_{n}(Y) & -\gamma_n(\sigma)
	\end{bmatrix} \begin{bmatrix}
	\xi_1 \\ \eta_1
	\end{bmatrix}=\begin{bmatrix}
	H_{n}(U) & 0 \\ H_{n}(Y) & -\gamma_n(\sigma)
	\end{bmatrix} \begin{bmatrix}
	\xi_2 \\ \eta_2
	\end{bmatrix}\!\!  \\ \!&\Rightarrow \eta_1=\eta_2.
	\end{split}%
	\end{equation}%
	Therefore, the data are informative for interpolation at $\sigma$ if and only if (\ref{Eqn_lin_0}) and (\ref{Eqn_uniqueness}) hold. Clearly, (\ref{Eqn_lin_0}) and (\ref{condition2}) are equivalent. It follows from Lemma~\ref{lemma_unique} in the appendix that (\ref{Eqn_uniqueness}) is equivalent to (\ref{condition3}). \qed
\end{pf}
\begin{rem}\label{remark_interpretation}
	The conditions (\ref{condition2}) and (\ref{condition3}) in Theorem~\ref{Thm_1} allow for an insightful interpretation. On the one hand, (\ref{condition3}) implies the existence of a system $\begin{bmatrix}
	q & -p
	\end{bmatrix} \in \Sigma_{U,Y}$ such that $P(\sigma) \neq 0$. On the other hand, (\ref{condition2}) guarantees that if there exists $\begin{bmatrix}
	q & -p
	\end{bmatrix} \in \Sigma_{U,Y}$ such that $P(\sigma)=0$, then $Q(\sigma)=0$.
\end{rem}
 An important consequence of Theorem~\ref{Thm_1} is that the data do not need to be informative for system identification, {i.e., \eqref{Cond_sysID}}, in order to be for interpolation. {This suggests to say that the informativity conditions for interpolation is weaker than those for identification.} Thus, it is possible that infinitely many systems explain the same data and they all have the same moment at a given interpolation point. 

When $\sigma \in \C \setminus \R$, it is natural to ask how data informativity for interpolation at $\sigma$ is related to informativity for its complex conjugate. The following proposition answers this question based on the results of Lemma~\ref{Thm_two_incl} and Theorem~\ref{Thm_1}.
\begin{prop}\label{propos_conj}
	The data $(U,Y)$ are informative for interpolation at $\sigma$ if and only if they are informative for interpolation at $\bar{\sigma}$ \footnote[2]{The notation $\bar{\sigma}$ denotes the complex conjugate of $\sigma$.}. Moreover, if the $0$-th moment at $\sigma$ is $M_0$, then the $0$-th moment at $\bar{\sigma}$ is $\bar{M}_0$.
\end{prop}
\vspace*{-6mm}
\begin{pf} Since the `if' part is evident, we prove only the `only if' part. Suppose that the data $(U,Y)$ are informative for interpolation at $\sigma$. In view of Theorem~\ref{Thm_1}, this means that (\ref{condition2}) and (\ref{condition3}) hold. By (\ref{condition2}), there exist $\xi \in \C^{T-n+1}$ and $M_0 \in \C$ such that (\ref{Eqn_lin2}) holds. By taking the complex conjugate of (\ref{Eqn_lin2}), we get
	\begin{equation}\label{Eqn_lincomplex}
	\begin{bmatrix}
	H_{n}(U) & 0 \\ H_{n}(Y) & -\gamma_n(\bar{\sigma})
	\end{bmatrix} \begin{bmatrix}
	\bar{\xi} \\ \bar{M}_0
	\end{bmatrix}=\begin{bmatrix}
	\gamma_n(\bar{\sigma}) \\ \!0
	\end{bmatrix},%
	\end{equation}
	where we have used that $\overline{\gamma_n(\sigma)}=\gamma_n(\bar{\sigma})$.
	This implies that
	\begin{equation}\label{condition2_conj}
	\rank\!\begin{bmatrix}
	H_{n}(U) &  \!0  & \!\gamma_n(\bar{\sigma}) \\ H_{n}(Y)  & \gamma_n(\bar{\sigma}) &  0
	\end{bmatrix}\!\!= \!\rank\!\begin{bmatrix}
	H_{n}(U) & 0\\ H_{n}(Y)  & \gamma_n(\bar{\sigma})
	\end{bmatrix}\!\!. \space\!\!%
	\end{equation}
	From (\ref{condition3}), we see that 
		\begin{equation}\label{condition3_conj}
	\rank\begin{bmatrix}
	H_{n}(U) & 0 \\ H_{n}(Y) & \gamma_n(\bar{\sigma})
	\end{bmatrix} = \rank\begin{bmatrix}
	H_{n}(U)  \\ H_{n}(Y) 
	\end{bmatrix} +1.%
	\end{equation}
	In view of Theorem~\ref{Thm_1}, (\ref{condition2_conj}) and (\ref{condition3_conj}) imply that the data $(U,Y)$ are informative for interpolation at $\bar{\sigma}$. From (\ref{Eqn_lincomplex}), we can conclude that $\bar{M}_0$ is the $0$-th moment at $\bar{\sigma}$.\qed
\end{pf}
\vspace*{-5mm}
To illustrate the first main result of this paper, we provide the following example.
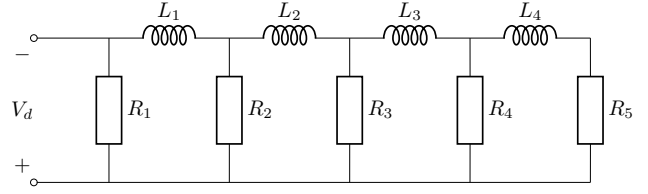
\begin{figure}[!t]
	\centering
	\captionsetup{justification=centering}
	\scalebox{0.8}{
		\begin{circuitikz}[scale=0.8]
			\node [ocirc](TW) at (-0.56,3) { }; \node [ocirc](TW) at (-0.56,0) {};
			\draw (-0.5,3)--(1,3) to [cute inductors, L=$L_1$] (3.5,3) to [cute inductors, L=$L_2$] (6,3) to [cute inductors, L=$L_3$] (8.5,3) to [cute inductors, L=$L_4$] (11,3);
			\draw (-0.5,0)-- (11,0);
			\draw (1,3) to [european resistors, R=$R_1$] (1,0);
			\draw (3.5,3) to [european resistors, R=$R_2$] (3.5,0);
			\draw (6,3) to [european resistors, R=$R_3$] (6,0);
			\draw (8.5,3) to [european resistors, R=$R_4$] (8.5,0);
			\draw (11,3) to [european resistors, R=$R_5$] (11,0);
			\node [] at (-0.8,0.35) {$+$}; \node [] at (-0.8,2.65) {$-$};
			\node[] at (-0.8,1.5) {$V_d$};
	\end{circuitikz}}
	\caption{RL circuit with four inductors and five resistors.}
	\label{Fig_RLcircuit}
\end{figure}
\begin{exmp}\label{example1}
	Consider the RL circuit depicted in Figure~\ref{Fig_RLcircuit}, which is a slight extension of \cite[Example~22]{jongsma2017model}. We take the currents through the inductors $L_1$, $L_2$, $L_3$ and $L_4$ as the states of the system, so $n=4$. The input is the voltage $V_d$. Finally, as the output, we take the current through the first inductor $L_1$. This leads to the continuous-time dynamical system
	\begin{eqnarray}
	\dot{x}&=&\begin{bmatrix}
	-\frac{R_2}{L_1} \!& \!\frac{R_2}{L_1} \!& \!0 \!& \!0 \\
	\frac{R_2}{L_2} \!& \!\frac{-(R_2+R_3)}{L_2} \!& \!\frac{R_3}{L_2} \!& \!0 \\
	0 \!& \!\frac{R_3}{L_3} \!& \!\frac{-(R_3+R_4)}{L_3} \!& \!\frac{R_4}{L_3} \\
	0 \!& \!0 \!&\! \frac{R_4}{L_4} \!& \frac{-(R_4+R_5)}{L_4} 
	\end{bmatrix} x + \begin{bmatrix}
	\frac{1}{L_1} \\0 \\0 \\0
	\end{bmatrix}u, \! \! \nonumber \\ \label{example_sys}
		y  &=&  \begin{bmatrix}
	1 &0 & 0 & 0
	\end{bmatrix}x.%
	\end{eqnarray}
	Let the inductances be given by $L_1=L_2=L_3=L_4=1 \ H$. For the resistors, we have $R_1=0.5\ \Omega$, $R_2=8\ \Omega$, $R_3=5\ \Omega$, $R_4=1\ \Omega$, $R_5=4\ \Omega$. 
	We stress that we assume that the exact model (\ref{example_sys}) is unknown and that only input-output data are available. {Specifically, suppose that the input data $U= \begin{bmatrix}
		\bar{u}_0 & \bar{u}_1 & \cdots & \bar{u}_T
		\end{bmatrix}$ are generated from an autonomous discrete-time system of the form ${\omega_{t+1}=S\omega_t,	\ \bar{u}_t=L\omega_t }$ with 
	{
	\begin{equation}
	S=\begin{bmatrix}
	{\sqrt{2}}& -1 & 0 & 0 \\ 1 & 0 & 0 & 0 \\ 0 & 0 & 1 & 1 \\ 0 & 0 & 0 & 1
	\end{bmatrix}, \ L= \begin{bmatrix}
	\frac{1}{2}  & \frac{1}{2} & \frac{1}{2} & \frac{1}{2}
	\end{bmatrix},\nonumber
	\end{equation}
initial condition
     ${\omega(0)=\begin{bmatrix}
    	1 & 0 & 0 & -0.1
    	\end{bmatrix}^T}$ and sampling period $\Delta= 0.2$ s.}
  The output data $Y= \begin{bmatrix}
  \bar{y}_0 & \bar{y}_1 & \cdots & \bar{y}_T
  \end{bmatrix}$ are the outputs of system \eqref{example_sys} with zero-order-hold input. We consider input-output data for $T=20$ ($4$ seconds). This leads to the samples 
	 depicted in Figure~\ref{Fig_plot_UY}.} {
	 Note that this method does not rely on steady-state data and/or zero initial condition in contrast to existing data-driven moment matching techniques, e.g., \cite{peherstorfer2017data} and \cite{scarciotti2017data}.}
	
	\begin{figure}[!t]
%
%
\definecolor{mycolor1}{rgb}{0.00000,0.44700,0.74100}%
\definecolor{mycolor2}{rgb}{0.85000,0.32500,0.09800}%
\begin{tikzpicture}{scale=0.5}

\begin{axis}[%
width=2.73in,
height=2.1in,
at={(0.898in,2.499in)},
xmin=0,
xmax=4,
xlabel style={font=\color{white!15!black}},
xlabel={Time},
ymin=-2.2,
ymax=1.3,
ylabel style={font=\color{white!15!black}},
axis background/.style={fill=white},
legend style={at={(0.680,0.795)}, anchor=south west, legend cell align=left, align=left, draw=white!15!black}
]
\addplot[const plot, color=mycolor1, line width=1.2pt] table[row sep=crcr] {%
0	0.45\\
0.2	1.10710678118655\\
0.4	1.05710678118655\\
0.6	0.300000000000001\\
0.8	-0.75\\
1	-1.50710678118655\\
1.2	-1.55710678118655\\
1.4	-0.900000000000001\\
1.6	0.0499999999999989\\
1.8	0.707106781186547\\
2	0.657106781186548\\
2.2	-0.0999999999999984\\
2.4	-1.15\\
2.6	-1.90710678118655\\
2.8	-1.95710678118655\\
3	-1.3\\
3.2	-0.350000000000002\\
3.4	0.307106781186547\\
3.6	0.257106781186549\\
3.8	-0.499999999999998\\
4	-1.55\\
};
\addlegendentry{Input}

\addplot[const plot, color=mycolor2, 
line width=1.2pt] table[row sep=crcr] {%
0	0\\
0.2	0.056260699290301\\
0.4	0.174227067634172\\
0.6	0.250922548387297\\
0.8	0.225264626789084\\
1	0.0981177320074527\\
1.2	-0.070766674426004\\
1.4	-0.198389498745495\\
1.6	-0.227262396347285\\
1.8	-0.159080778955464\\
2	-0.0536870389743825\\
2.2	0.00603889089496142\\
2.4	-0.0372141802861863\\
2.6	-0.181561397153922\\
2.8	-0.36697286455205\\
3	-0.510387545666921\\
3.2	-0.554320822391539\\
3.4	-0.500492004651978\\
3.6	-0.40877316004894\\
3.8	-0.362074928697628\\
4	-0.417738640644862\\
};
\addlegendentry{Output}

\end{axis}

\end{tikzpicture}%
		\centering
		\caption{Input/output data with sampling period $\Delta=0.2$~s.}
				\captionsetup{justification=centering}
		\label{Fig_plot_UY}
	\end{figure}
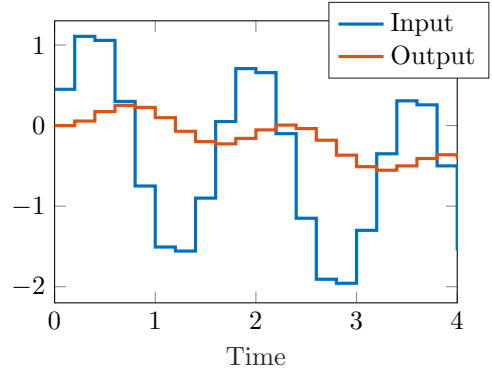
	%
	%
	
	It can be verified that condition (\ref{Cond_sysID}) does not hold for these data. As such, the data are not informative for system identification. Instead, there are (infinitely) many systems of the form (\ref{Eqn_datasys}) with order $n=4$ that explain the data. 
{{Suppose that we aim at interpolation at 
	$\sigma_1=1$, $\sigma_{2,3}=\frac{1}{\sqrt{2}}\pm\frac{i}{\sqrt{2}}$ and $\sigma_{4,5}=\pm i$, which correspond to frequencies $0$ rad/s, $ \frac{5 \pi}{4}$ rad/s and $ \frac{5 \pi}{2}$ rad/s, respectively,  as $\sigma_1=e^{0 \cdot \Delta}$, $\sigma_{2,3}=e^{\pm (5 \pi i \Delta)/{4}}$ and $\sigma_{4,5}=e^{\pm (5 \pi i \Delta)/{2}}$. 
	One can verify by Theorem~\ref{Thm_1} that the data $(U,Y)$ are informative for interpolation at $\sigma_1$, $\sigma_2$ and $\sigma_3$. Similarly, it can be verified that the data are {\em not\/} informative for interpolation at $\sigma_{4,5}$. Finally, the moments of order $0$ at $\sigma_1$, $\sigma_2$ and $\sigma_{3}$ are given by $1.575$, $0.0031 - 0.1417i$ and $0.0031 +0.1417i$, respectively, which are obtained by solving \eqref{Eqn_lin2}}.
}
\end{exmp}

\section{Data informativity for higher-order  moment matching}\label{Section_higher}
We recall that higher-order moments are defined recursively, see Definition~\ref{Def_kthmoment}. Then, the following results provide a counterpart of Lemma~\ref{Thm_two_incl} and Theorem~\ref{Thm_1} for higher-order moments.
%
\begin{lem}\label{Lem_incl_Mk}
	Let $\sigma$ and $M_j$ for $j=0,1,\ldots,k$ be complex numbers. Then, the inclusion $\Sigma_{U,Y} \subseteq \Sigma_{\sigma,M_k}^k$ holds if and only if there exists $\xi \in \C^{T-n+1}$ such that
		\begin{equation}\label{Eqn_linxiMk}
	\ \begin{bmatrix}
	H_{n}(U)  & 0 \\ H_{n}(Y) &  -\gamma_n(\sigma) 
	\end{bmatrix} \!\! \begin{bmatrix}
	\xi \\ M_k
	\end{bmatrix}\!=\!\begin{bmatrix}
	\gamma^{(k)}_n(\sigma) \\ \sum_{j=0}^{k-1} \binom{k}{j}M_j\gamma^{(k-j)}_n(\sigma)
	\end{bmatrix}\ \!\!\!\!.\!\!\!\!\!\!
	\end{equation}
	\end{lem}
	\vspace*{-8mm}
\begin{pf}
	Note that $\Sigma_{U,Y} \subseteq \Sigma_{\sigma,M_k}^k$ is equivalent to saying that every solution of 
	(\ref{linear1}) is also a solution of (\ref{Eqn_kthmoment}). The rest of the proof is similar to that of Lemma~\ref{Thm_two_incl}, hence it is omitted. \qed
	\end{pf}
	\vspace*{-5mm}
Using Lemma~\ref{Lem_incl_Mk}, the following result provides necessary and sufficient conditions for data informativity for higher-order moment matching. 
\begin{thm}\label{thm_Mk}
	Let $k>0$ and suppose that the data $(U,Y)$ are informative for moment matching of order $j$ at $\sigma$ for all $j$ with $0 \leq j<k$. Let  $M_j$ denote the corresponding moments. Then, the data $(U,Y)$ are informative for moment matching of order $k$ at $\sigma$ if and only if
	\begin{equation}\label{conditionkthmoment}
	\begin{split}
	\rank&\begin{bmatrix}
	H_{n}(U) & 0 & \gamma^{(k)}_n(\sigma) \\ H_{n}(Y) & \gamma_n(\sigma) & \sum_{j=0}^{k-1} \binom{k}{j}M_j\gamma^{(k-j)}_n(\sigma)
	\end{bmatrix} \\
	&=\rank\begin{bmatrix}
	H_{n}(U) & 0 \\ H_{n}(Y) & \gamma_n(\sigma)
	\end{bmatrix}.%
	\end{split}%
	\end{equation}%
\end{thm}%
\begin{pf}
The assumption given in this theorem that the data are informative for moment matching of order $j$ at $\sigma$ for all $j$ with $0 \leq j<k$ satisfies the first condition of Definition~\ref{Def_informativityMk}. In addition, the data are informative for moment matching of order $k$ at $\sigma$ if there exists a unique $M_k$ such that $\Sigma_{U,Y} \subseteq \Sigma^k_{\sigma,M_k}$. The existence of such an $M_k$ is equivalent to the existence of $\xi$ such that
		\begin{equation}\label{Eqn_linxiMk_inproof}
	\ \begin{bmatrix}
	H_{n}(U)  & 0 \\ H_{n}(Y) &  -\gamma_n(\sigma) 
	\end{bmatrix} \!\! \begin{bmatrix}
	\xi \\ M_k
	\end{bmatrix}\!=\!\begin{bmatrix}
	\gamma^{(k)}_n(\sigma) \\ \sum_{j=0}^{k-1} \binom{k}{j}M_j\gamma^{(k-j)}_n(\sigma)
	\end{bmatrix}\!\!\!\!\!\!
	\end{equation}
	due to Lemma~\ref{Lem_incl_Mk}. Clearly, this is equivalent to (\ref{conditionkthmoment}). 
	
	Note that the right hand side of (\ref{Eqn_linxiMk_inproof}) does not play a role in determining the uniqueness of $M_k$. Since the data have been assumed to be informative for interpolation at $\sigma$, condition (\ref{condition3}) in Theorem~\ref{Thm_1} holds. This guarantees the uniqueness of $M_k$ since the matrix in the left hand side of (\ref{Eqn_linxiMk_inproof}) is the same to that of (\ref{Eqn_lin_0}). \qed
\end{pf}
Analogous to Theorem~\ref{Thm_1}, this theorem provides conditions under which all systems explaining the data have the same higher-order moment at a given interpolation point.

The conjugacy relation between $0$-th moments at conjugate pairs of interpolation points can be proven in a similar way to the proof of Proposition~\ref{propos_conj}.
\begin{prop}\label{propos_conj_kth}
	Let $k>0$. The data $(U,Y)$ are informative for moment matching of order $k$ at $\sigma$ if and only if they are informative for moment matching of order $k$ at $\bar{\sigma}$. Moreover, if the $k$-th moment at $\sigma$ is $M_k$, then the $k$-th moment at $\bar{\sigma}$ is $\bar{M}_k$. %
\end{prop}

We illustrate the results of Theorem~\ref{thm_Mk} by means of an example.
{
\begin{exmp}\label{example2}
	Consider the system and input-output data studied in Example~\ref{example1}. It can be checked that for $\sigma=1$, conditions \eqref{conditionkthmoment} holds for $k=1$.
	Hence, the data are informative for moment matching of order $1$ at $\sigma=1$. By solving linear equation
	(\ref{Eqn_linxiMk}) (for $k=1$), we obtain 
	$M_1=-31.8437$.
\end{exmp}
}

So far, our discussion considered a single interpolation point $\sigma$ and its desired order of moment $k$. Let a collection of pairs of interpolation points and their desired order of moments 
\begin{equation}\label{e:PP}
\PP= \left\{ \left. (\sigma_i,k_i) \ \right| \ i=1,2,\ldots,s \right\}
\end{equation}
be given. In view of Propositions~\ref{propos_conj} and \ref{propos_conj_kth}, we assume that {$(\bar{\sigma}_i,k_i)  \in \PP$ whenever $(\sigma_i,k_i) \in \PP$}. By applying Theorems~\ref{Thm_1} and \ref{thm_Mk}, one can verify whether the data are informative for moment matching for each pair. If so, (\ref{Eqn_lin2}) and (\ref{Eqn_linxiMk}) result in the corresponding moments
\begin{equation}\label{e:MM}
\MM_i = \left\{\left. (\sigma_i,M_j^i) \ \right| \ j=0,1,\ldots,k_i \right\},
\end{equation}
where $M_j^i$ denotes the $j$-th moment at $\sigma_i$.



\section{Reduced-order model by data-driven moment matching}\label{Section_rom}
In this section, we will investigate how reduced-order models can be computed from data that are informative for moment matching. As the results of Theorems~\ref{Thm_1}~and~\ref{thm_Mk} lead to the computation of moments at given interpolation points, obtaining such reduced-order model is essentially a rational interpolation problem.

Let a reduced-order model $\hat{\Sigma}$ of order $r$ be given by
\begin{equation}\label{TF_red}
\!\begin{split}
&\left(z^r+\hat{p}_{r-1}z^{r-1}+\cdots+\hat{p}_1z+\hat{p}_0\right)y_t \\ &  \quad \quad \quad = \left(\hat{q}_rz^r+\hat{q}_{r-1}z^{r-1}+\cdots+\hat{q}_1z+\hat{q}_0\right)u_t.\!
\end{split}\!
\end{equation}
As before, we collect the parameters of \eqref{TF_red} as 
$\hat{p}=\begin{bmatrix} \hat{p}_0 & \hat{p}_1 & \cdots & \hat{p}_{r-1} \end{bmatrix}$ and $\hat{q}=\begin{bmatrix}
\hat{q}_0 \ \hat{q}_1 \ \cdots \ \hat{q}_r
\end{bmatrix}$. Denote by  $\hat{P}(z)$ and $\hat{Q}(z)$ the polynomials on the left- and right-hand side of (\ref{TF_red}), respectively. Then, the model (\ref{TF_red}) interpolates or matches the $0$-th moment at $\sigma$ if $M_0$ in (\ref{Eqn_lin2}) satisfies $\hat{Q}(\sigma)=M_0\hat{P}(\sigma)$ which is equivalent to
\begin{equation*}
\begin{bmatrix}
\hat{q} & -\hat{p}
\end{bmatrix}\begin{bmatrix}
\gamma_r(\sigma) \\ M_0{\gamma}_{r-1}(\sigma) 
\end{bmatrix}=M_0\sigma^r.
\end{equation*}
More generally, for $\PP$ and $\MM_i$ as in \eqref{e:PP} and \eqref{e:MM}, a reduced-order model parameterization by $\begin{bmatrix}
\hat{q} & -\hat{p}
\end{bmatrix}$ must satisfy the linear equations
\begin{equation}\label{Eqn_full_intp}
\begin{bmatrix}
\hat{q} & - \hat{p}
\end{bmatrix}\begin{bmatrix}
\Gamma^r(\sigma_i)\\
\Gamma^r_{M}(\sigma_i)
\end{bmatrix}_{1:(2r+1)} 
=\begin{bmatrix}
\Gamma^r(\sigma_i)\\
\Gamma^r_{M}(\sigma_i)
\end{bmatrix}_{(2r+2)}
\end{equation}
for $i=1,2,\ldots,s$, where
 	\begin{equation*}
\!\begin{pmat}[{..}]
\!\Gamma^r(\sigma_i) \cr\-
\!\Gamma^r_{M}(\sigma_i) \!\cr 
\end{pmat}\!\!=\!\!\begin{pmat}[{..}]
\gamma_r(\sigma_i) &  \cdots\! & \gamma_r^{(k_i)}(\sigma_i) \cr\-
M_0^i\gamma_r(\sigma_i) & \cdots & \sum_{j=0}^{k_i} \binom{
	k_i}{ j}M_j^i\gamma^{(k_i-j)}_r(\sigma_i) \cr
\end{pmat}\!\!.
\end{equation*}
Here, the notation $X_{1:(2r+1)}$  and $X_{(2r+2)}$ in (\ref{Eqn_full_intp}) denote the first $(2r+1)$ rows and $(2r+2)$-th row of a matrix $X$, respectively.
	Note that since $\begin{bmatrix}
		\hat{q} & - \hat{p}
	\end{bmatrix}$ is restricted to be real, then if $\sigma_i$  is complex, we split \eqref{Eqn_full_intp} 
	into its real and imaginary parts.

Now, we define
$\hat{\Sigma}_{r,\PP} = \{\begin{bmatrix}
\hat{q} & -\hat{p}
\end{bmatrix} \in \R^{1 \times (2r+1)}\ | \ (\ref{Eqn_full_intp}) \ \text{holds}\}.$
Clearly, $\hat{\Sigma}_{r,\PP}$ consists of all models of order $r$ matching the moments $M_j^i$ at the interpolation point $\sigma_i$ for all $i=1,2,\ldots,s$ and $j=0,1,2,\ldots,k_i$. The following theorem readily follows from the solvability conditions of the linear equation given in (\ref{Eqn_full_intp}).
\begin{thm}\label{thm_r}
	Given the data of interpolation points and moments $\MM_i$ for $i=1,2,\ldots,s$ as in \eqref{e:MM}. Then, $\hat{\Sigma}_{r,\PP} \neq \emptyset$ if and only if
	\begin{equation}\label{rankcondr}
	\begin{split}
	\rank& \begin{bmatrix}
	\Gamma^r(\sigma_1) & \Gamma^r(\sigma_2) & \cdots & \Gamma^r(\sigma_s)\\
	\Gamma^r_{M}(\sigma_1) & \Gamma^r_{M}(\sigma_2) & \cdots & \Gamma^r_{M}(\sigma_s)
	\end{bmatrix}_{1:(2r+1)}\quad \quad \\
	&  \quad   =\rank\begin{bmatrix}
	\Gamma^r(\sigma_1) & \Gamma^r(\sigma_2) & \cdots & \Gamma^r(\sigma_s)\\
	\Gamma^r_{M}(\sigma_1) & \Gamma^r_{M}(\sigma_2) & \cdots & \Gamma^r_{M}(\sigma_s)
	\end{bmatrix}.
	\end{split}
	\end{equation}
\end{thm}
	We note that in this paper, we do not restrict $\begin{bmatrix}
		\hat{q} & -\hat{p}
	\end{bmatrix} \in \hat{\Sigma}_{r,\PP}$ to be minimal, i.e., the polynomials $\hat{P}(z)$ and $\hat{Q}(z)$ might not be coprime. 	In addition, it is also obvious that if $r\geq k^*-1$ where $k^*=\sum_{i=1}^{s} (k_i +1)$, then \eqref{rankcondr} always holds.


The following example illustrates the computation of the reduced-order model.

\begin{figure}[!t]
	\input{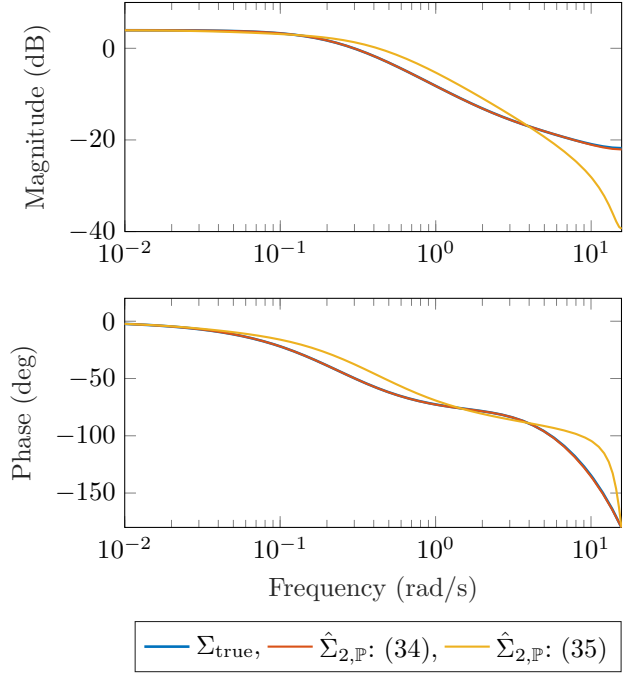}
	\centering
	\caption{Comparison of the Bode plot of a second-order to that of the higher-order model.}
	\captionsetup{justification=centering}
	\label{Fig_bode_ori_red}
\end{figure}

\begin{exmp}\label{example_rom}
{
	From Examples \ref{example1} and \ref{example2}, we have $\PP=\{(\sigma_1,1),(\sigma_2,0),(\sigma_3,0)\}$ and $\MM_i$ as follows:
	\begin{equation*}
	\begin{split}
	\MM_1&\!=\!\left\{(\sigma_1,M_0^1),(\sigma_1,M_1^1)\right\}\!=\!\left\{(1,1.575),(1,-31.8437)\right\}\!,\\
	\MM_2&\!=\!\left\{(\sigma_2,M_0^2)\right\}\!=\!\big\{\big({1}/{\sqrt{2}}+{i}/{\sqrt{2}},0.0031 - 0.1417i\big)\big\},\\
	\MM_3&\!=\!\left\{(\sigma_3,M_0^3)\right\}\!=\!\big\{\big({1}/{\sqrt{2}}-{i}/{\sqrt{2}},0.0031 + 0.1417i\big)\big\}.
	\end{split}
	\end{equation*}
It can be checked that condition (\ref{rankcondr}) does not hold for $r=1$.
	Therefore, $\hat{\Sigma}_{1,\PP}=\emptyset$.
	Meanwhile, it is satisfied with $r=2$. 	Hence, $\hat{\Sigma}_{2,\PP} \neq \emptyset$.} 

We recall that \eqref{Eqn_full_intp} characterizes all reduced-order systems that achieve moment matching (i.e., the set $\hat{\Sigma}_{2,\PP}$), of which two examples are written in the form (\ref{TF_red}) as
{
	\begin{equation}\label{sys_example}
	\begin{split}
	(z^2 - &1.191 z + 0.2268) y_t \\ &\quad = (0.0001953 z^2 + 0.1236 z - 0.06692)u_t
	\end{split}
	\end{equation}
	and
		\begin{equation}\label{sys_example2}
	\begin{split}
	(z^2 - &1.911 z + 0.9116) y_t \\ &\quad = (0.04622 z^2 + 0.02116 z - 0.06605)u_t.
	\end{split}
	\end{equation}}
\end{exmp}
The Bode plot of the reduced-order models \eqref{sys_example} and \eqref{sys_example2} compared to the higher-order model of Example~\ref{example1} are given in Figure~\ref{Fig_bode_ori_red}. Here, of course, we do not identify the higher-order model. Figure~\ref{Fig_bode_ori_red} shows that the reduced-order model \eqref{sys_example} captures the magnitude and phase of the higher-order model well, but \eqref{sys_example2}  does not. {Nevertheless, all of the curves are intersected at frequency $0$ rad/s and $\frac{5\pi}{4}$ rad/s}. This indicates that indeed all systems in $\hat{\Sigma}_{2,\PP}$ achieve the desired moment matching, but the optimality on approximating the higher-order model is another issue. We stress that this is a well-known feature of moment matching methods and not specific to our data-driven approach.

Often reduced-order models are expected to preserve certain properties of the original models. Stability is one of the most common as well important property to be preserved. Next, we investigate conditions under which one can choose a stable reduced-order model. Note that stability of a reduced-order model $\begin{bmatrix}
\hat{q} & -\hat{p}
\end{bmatrix} \in \hat{\Sigma}_{r,\PP}$ is purely determined by $\hat{p}$, {i.e., one can choose $\hat{p}$ such that the roots of its corresponding polynomial are in the unit disc}. Motivated by this observation, we provide a sufficient condition such that $\hat{p}$ can be chosen arbitrarily while $\begin{bmatrix}
\hat{q} & -\hat{p}
\end{bmatrix} \in \hat{\Sigma}_{r,\PP}$. 

\begin{thm}
	Given $\PP$ and $\MM_i$ as in \eqref{e:PP} and (\ref{e:MM}), respectively and let $k^*=\sum_{i=1}^{s} (k_i +1)$. If $r \geq k^*-1$ then for every $\hat{p} \in \R^{1 \times r}$ there exists $\hat{q} \in \R^{1 \times (r+1)}$ such that $\begin{bmatrix}
	\hat{q} & -\hat{p}
	\end{bmatrix} \in \hat{\Sigma}_{r,\PP}$.
\end{thm}
\begin{pf}	
   	We know that $\begin{bmatrix}
   \hat{q} & -\hat{p}
   \end{bmatrix} \in \hat{\Sigma}_{r,\PP}$ if and only if 
   \begin{equation}\label{e:rowspaceG1G2}
   \begin{bmatrix}
   \hat{q} & -\hat{p} & -1
   \end{bmatrix}\begin{bmatrix}
   \Gamma_1\\
   \Gamma_2
   \end{bmatrix}=0,
   \end{equation}
   where $\Gamma_1=\begin{bmatrix}
   \Gamma^r(\sigma_1) & \Gamma^r(\sigma_2) & \cdots & \Gamma^r(\sigma_s)
   \end{bmatrix}, \ \text{and }
   \Gamma_2=\begin{bmatrix}
   \Gamma^r_{M}(\sigma_1) & \Gamma^r_{M}(\sigma_2) & \cdots & \Gamma^r_{M}(\sigma_s)
   \end{bmatrix}$.
  	It is clear that $r\geq k^*-1$ guarantees $\hat{\Sigma}_{r,\PP} \neq \emptyset$.
   Particularly, since the structure of $\Gamma_1$ is a Vandermonde matrix, then $\rank(\Gamma_1)=k^*$ which implies \eqref{rankcondr}. Equation \eqref{e:rowspaceG1G2} also means that for every $\hat{p}$ there exists $\hat{q}$ such that $\begin{bmatrix}
	\hat{q} & -\hat{p}
	\end{bmatrix} \in \hat{\Sigma}_{r,\PP}$ if and only if 
	\begin{equation}\label{e:inclrowspaces}
 	\rowspace(\Gamma_2) \subseteq \rowspace(\Gamma_1).
	\end{equation}
    Since we know that $\rank(\Gamma_1)=k^*$, i.e., $\rowspace(\Gamma_1)=\R^{1 \times k^*}$, then indeed \eqref{e:inclrowspaces} holds. This finalizes the proof. \qed
\end{pf}
{
\begin{exmp}
	 Consider $\PP$ as in Example~\ref{example_rom}. 
	 Let us take $r=3$, in which case \eqref{rankcondr} is guaranteed.
	Suppose that we desire to place the poles at $\{0.25,0.4,0.95\}$ to guarantee stability of the resulting reduced-order model. This corresponds to the choice $\hat{p}=\begin{bmatrix}
	 -0.095 &
	0.7175 &
	-1.6
	\end{bmatrix}$.  Next, we can compute $\hat{q}$ by solving \eqref{Eqn_full_intp} with given $\hat{p}$ to obtain a reduced-order model $ (z^3 - 1.6 z^2 + 0.7175 z - 0.095)y_t =(-0.05625 z^3 + 0.2624 z^2 - 0.2574 z + 0.08674)u_t$.
\end{exmp}
}
\section{Conclusion}\label{Section_conclusion}

Motivated by data-driven model reduction, the concept of data informativity for moment matching using input-output data is proposed in this paper. The main results are necessary and sufficient conditions based on the rank of the Hankel matrix of the data and given (complex) interpolation points for computing the moments at these points. These conditions are \emph{weaker} than those for system identification, which implies that the data-driven approach for moment matching can be performed on data that are not informative for system identification. Namely, the precise value of moments can be extracted even though the high-order system model is not available. 
The moments computed from the informativity framework are then exploited to construct a reduced-order model by solving a rational interpolation problem. Moreover, the condition to enforce prescribed poles upon the reduced-order model is discussed. 

This work shows the advantages of using the informativity framework for data-driven model reduction by moment matching. We are confident that this gives the way to solve other data-driven model reduction problems, e.g., with preserving specific system properties {and selection of interpolation points to minimize the error.} This study forms one of the issues of our future research.

\begin{ack}                               
This paper is based on research developed in the DSSC Doctoral Training Programme, co-funded through a Marie Skłodowska-Curie COFUND (DSSC 754315).  
\end{ack}


\bibliographystyle{plain}        
\bibliography{autosam}           



\appendix
\section{Auxiliary results}   \label{appendixA} 
For the sake of completeness, we present the following two elementary linear algebra results that are employed in proving the main results. {The proofs are elementary and hence omitted.}
\begin{lemapp}\label{Lemma2tr}
	Let $A_i \in \R^{n \times m_i}$ and $b_i \in \R^{1 \times m_i}$ for $i=1,2$. Consider the sets $\mathcal{X}_i=\left\lbrace \xi \in \R^{1 \times n} \ | \ \xi A_i=b_i \right\rbrace$ and assume that $\mathcal{X}_1$ is nonempty. Then, $\mathcal{X}_1 \subseteq \mathcal{X}_2$ if and only if
	\begin{equation}\label{incl_X1X2}
	\leftker\begin{bmatrix}
	A_1 \\ b_1
	\end{bmatrix} \subseteq \leftker\begin{bmatrix}
	A_2 \\ b_2
	\end{bmatrix}.
	\end{equation}
\end{lemapp}
\begin{lemapp}\label{lemma_unique}
	Let $a \in \C^n$ and $A \in \C^{n \times n}$. Then, the following statements are equivalent:
	\begin{enumerate}
		\item[(i)] 	If 
		$\begin{bmatrix}
		A & a
		\end{bmatrix}\begin{bmatrix}
		\xi_1 \\ \eta_1
		\end{bmatrix}= \begin{bmatrix}
		A & a
		\end{bmatrix}\begin{bmatrix}
		\xi_2 \\ \eta_2
		\end{bmatrix}$,
		then $\eta_1 = \eta_2$.
		\item[(ii)] 
		$\rank\begin{bmatrix}
		A & a
		\end{bmatrix} = \rank
		A + 1$.
	\end{enumerate}
\end{lemapp}

\end{document}